\documentclass[twocolumn,showpacs,preprintnumbers,amsmath,amssymb,footinbib]{revtex4-1}
\usepackage[utf8]{inputenc}
\usepackage[english]{babel}
\usepackage{lmodern}
\usepackage{amsmath}
\usepackage{graphicx}
\usepackage{amssymb}
\usepackage{epstopdf}
\usepackage{ upgreek }
\usepackage{color}
\usepackage{setspace}
\usepackage{mathtools}

\graphicspath{{Figures/}}

\usepackage[toc,page]{appendix}
\usepackage{tocloft}

\begin{document}

\title[]{Probing the spectrum of the Jaynes--Cummings--Rabi model by its isomorphism to an atom inside a parametric amplifier cavity}

\author{R.~Guti\'{e}rrez-J\'{a}uregui}
\email[Email:]{r.gutierrez.jauregui@gmail.com}
\affiliation{Department of Physics, Columbia University, New York, NY, USA.}
\author{G.~S.~Agarwal}
\email[Email:]{girish.agarwal@tamu.edu}
\affiliation{Institute for Quantum Science and Engineering, and Departments of Biological and Agricultural Engineering, and Physics and Astronomy Texas A \& M University, College Station, TX 77843}

\date{}
\pacs{}

\begin{abstract}
We show how the Jaynes--Cummings--Rabi model of cavity quantum electrodynamics can be realized via an isomorphism to the Hamiltonian of a qubit inside a parametric amplifier cavity. This realization clears the way to observe the full spectrum of the Rabi model via a probe applied to a parametric amplifier cavity containing a qubit and a parametric oscillator operating below threshold. An important outcome of the isomorphism is that the actual frequencies are replaced by detunings which make it feasible to reach the ultra-strong coupling regime. We find that inside this regime the probed spectrum displays a narrow resonance peak that is traced back to the transition between ground and first excited states. The exact form of these states is given at an energy crossing and then extended numerically. At the crossing, the eigenstates are entangled states of field and atom where the field is found inside squeezed cat states.
\end{abstract}

\maketitle

%%%%%%%%%%%%%%%%%%%%%%%%%%%%%%%%%%%%%%%%%%%%%%%%%%%%%%%%%%%%%%%%%%%%%%%%%%%%%%%%%%%%%%%%%%%%%%%%%%%%%%
\section{Introduction}
%%%%%%%%%%%%%%%%%%%%%%%%%%%%%%%%%%%%%%%%%%%%%%%%%%%%%%%%%%%%%%%%%%%%%%%%%%%%%%%%%%%%%%%%%%%%%%%%%%%%%%

It is well-known that a weak harmonic field is particularly efficient to induce a transition between two states of an atom when it oscillates at a frequency close to the Bohr frequency separating the states. The field can be used in this way to probe the atomic energy spectrum and provide a window into the underlying processes that rule the atomic dynamics. As the intensity of the field is ramped up, it perturbs the atomic response in a way that signals that field and atom have coupled into a single system. The energy diagram of this composite system displays a rich structure with crossings and avoided-crossings whose locations are given by an interplay of atomic transition frequency, field frequency, and coupling strength~\cite{Cohen_1998}. This diagram carries information of an idealized light-matter coupling and has been the subject of extensive research ranging from the many-photon~\cite{Bloch_1940,Shirley_1965,Pegg_1970,Stenholm_1972,Cohen-Tannoudji_1973a} down to the single-photon limit~\cite{Cohen-Tannoudji_1973b,Haroche_2016,Girvin_2011,Leibfried_2003}. 

Recent advances have made it possible to increase the single-photon coupling to about 10\% of the atomic transition frequency~\cite{Ciuti_2005,Anappara_2009,Forn_Diaz_2010,Niemczyk_2010,Pedernales_2015,Kraglund_2017,Forn-Diaz_2019,Kockum_2019}. Due to this strong coupling, models used to describe these systems need to move beyond common simplifications, as the rotating wave approximation. And, in the idealized case of a two-state atom coupled to a single mode of the field under the dipole approximation, the system is accurately described by the Hamiltonian ($\hbar = 1$)
\begin{align} \label{eq:Rabi_Hamiltonian}
\hat{\mathcal{H}}_{\eta} = \omega_{c} \hat{a}^{\dagger}\hat{a} +  \omega_{a} \hat{\sigma}_{+}\hat{\sigma}_{-} + &\lambda^{\prime} (\hat{a} \hat{\sigma}_{+} + \hat{a}^{\dagger} \hat{\sigma}_{-}) \nonumber \\
+ \eta &\lambda^{\prime} (\hat{a} \hat{\sigma}_{-} + \hat{a}^{\dagger} \hat{\sigma}_{+})  \, 
\end{align}
with $\eta = 1$. Here, $\hat{a}$ and $\hat{a}^{\dagger}$ are annihilation and creation operators for the field mode; $\hat{\sigma}_{+}$ and $\hat{\sigma}_{-}$ raising and lowering operators for the atomic levels; and $\omega_{c}$ ($\omega_{a}$), $\lambda$ are parameters that denote the mode (atomic transition) frequency and coupling strength. The inclusion of $\eta$ makes it possible to create a bridge between the Jaynes-Cummings limit with $\eta = 0$ (suitable for $\omega_{c} \gg \lambda$) and the Rabi limit with $\eta = 1$. This parameter can then be varied to study the effects of an increasing coupling strength. 

The energy spectrum of this model---refered to as the Jaynes--Cummings--Rabi (JC--Rabi) model---was recently obtained by Tomka and collaborators~\cite{Tomka_2014,Tomka_2015} whose analytic results extend on the seminal work by Braak~\cite{Braak_2011} for the Rabi limit. The theoretical advances raise the natural question on how to design systems where this spectrum can be probed. Current approaches have moved to a driven-dissipative setting where atomic and mode frequencies are changed into detunings to a driving field~\cite{Ballester_2012} and coupling terms are generated from interactions between additional levels~\cite{Dimer_2007} or degrees of freedom~\cite{Morales_2018}. Here, we propose a different method to study the JC--Rabi model where we present a realizable system whose model Hamiltonian is isomorphic to Eq.~(\ref{eq:Rabi_Hamiltonian}). In this way, the system can be used to probe the energy spectrum of the JC-Rabi model over a large range of parameters. 

\begin{figure}[ht]
\begin{center}
\includegraphics[width=.485\linewidth]{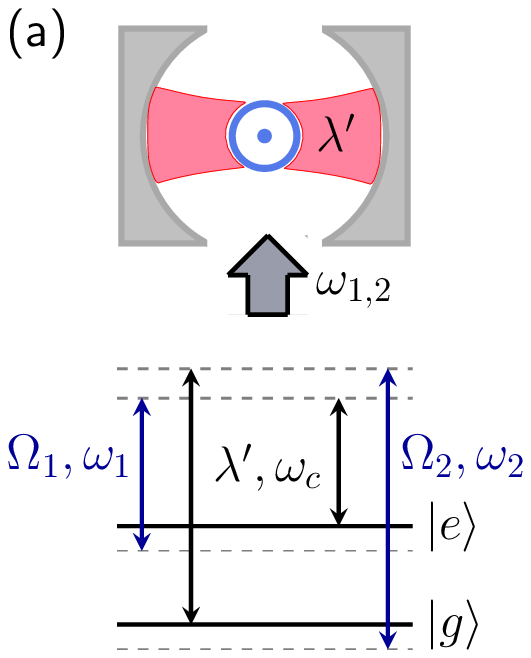}  \includegraphics[width=.485\linewidth]{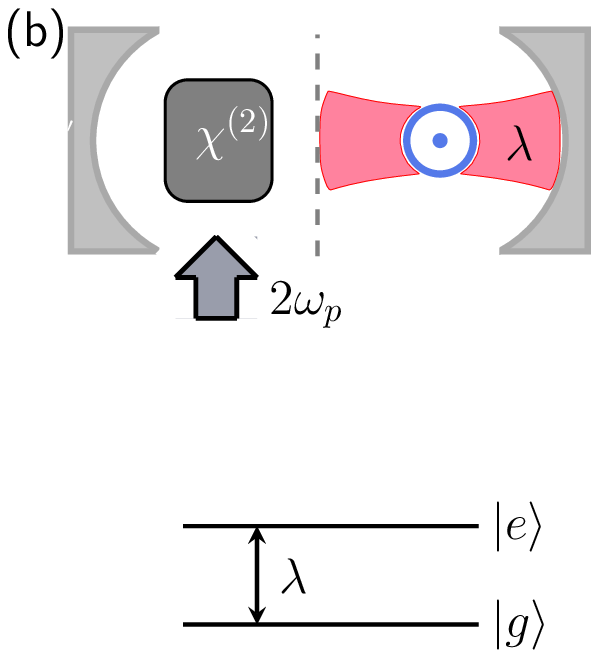} 
\caption{Sketch of two optical amplification processes that generate the JC-Rabi model from the JC model. (a) Coupled lambda transitions introduced in Ref.~\cite{Dimer_2007} . (b) Qubit inside a parametric amplifier discussed in Sec.~\ref{sec:background}}\label{fig:sketch}
\end{center}
\end{figure}

In Figure~\ref{fig:sketch} we sketch two possible realizations of the JC--Rabi model that are relevant to this work. On the left panel the model is generated from two metastable-states coupled through a pair of lambda transitions as originally proposed in Ref.~\cite{Dimer_2007}. This proposal has found great success on atomic gases where the necessary level structure is encountered~\cite{Kroeze_2018,Morales_2018,Gutierrez_2018}. On the right panel an isomorphic Hamiltonian is generated from a two-state system coupled to a parametric oscillator. The basic ingredients being available in superconducting circuits architectures~\cite{Siddiqi_2013,Siddiqi_2016}. The isomorphism is presented below where it is shown that the conditions on the level structure are relaxed in exchange of independent control of the parameters.

The advantage of using a cavity with parametric amplifier was recognized in the context of optomechanics~\cite{Huang_2009} where it was shown how the parametric coupling can result in normal mode splitting and the squeezing of the mechanical oscillator~\cite{Agarwal_2016}. More recently the use of parametric amplifier has been advocated in enhancing collective effects~\cite{Qin_2018}  and in quantum phase transitions~\cite{Agarwal_2020}. The idea to reach the strong coupling regime using a two-photon drive was also presented recently in Ref.~\cite{Leroux_2018} where the authors consider an adiabatic switch on the drive to reach the ground state of a ultra-strong coupled system. 

%%%%%%%%%%%%%%%%%%%%%%%%%%%%%%%%%%%%%%%%%%%%%%%%%%%%%%

The outline of the paper is as follows. In Sec.~\ref{sec:background} we introduce a model to describe a qubit inside a parametric amplifier operating below threshold and show its equivalence to the JC--Rabi model. The eigenvalues and eigenstates of the parametric model are calculated. In Sec.~\ref{sec:spectrum} we simulate the absorption spectrum of this system as probed by a weak coherent field. The spectrum displays several peaks that are connected to the eigenstates obtained before. We find a sharp peak whose breath decreases as we dwelve deeper into the strong coupling regime. Section~\ref{sec:outside} is used to describe the eigenstates that lead to this peak. The states are shown to form a long-living pair described by two-photon cat states conditioned to the state of the qubit. Sections~\ref{sec:discussion} and~\ref{sec:conclusion} are left for discussion and concluding remarks.
%%%%%%%%%%%%%%%%%%%%%%%%%%%%%%%%%%%%%%%%%%%%%%%%%%%%%%%%%%%%%%%%%%%%%%%%%%%%%%%%%%%%%%%%%%%%%%%%%%%%%%
\section{JC--Rabi model and the qubit inside a parametric amplifier}\label{sec:background}
%%%%%%%%%%%%%%%%%%%%%%%%%%%%%%%%%%%%%%%%%%%%%%%%%%%%%%%%%%%%%%%%%%%%%%%%%%%%%%%%%%%%%%%%%%%%%%%%%%%%%%

In the proposed realization a qubit is placed inside a parametric cavity as sketched in Figure~\ref{fig:sketch}b. The cavity supports two modes, labeled pump and subharmonic, that couple to one another through a nonlinear material of second-order susceptibility $\chi^{(2)}$. We consider the pump mode to be highly populated and treat it as a classical field of constant amplitude $G/2$ and frequency $\omega_{p}$ that drives the subharmonic mode via a two-photon process~\cite{Carmichael_2008}. In addition, the subharmonic mode is coupled to a qubit with coupling strength $\lambda$ under conditions that allow for the dipole and rotating-wave approximations. 
%The now driven subharmonic mode is then coupled to a qubit with coupling strength $\lambda$ under conditions that allow for the dipole and rotating-wave approximations. 

The master equation for the density operator of the qubit-mode system $\rho$ is 
\begin{equation}\label{eq:master_eq}
\dot{\rho} = -i \left[ \hat{\mathcal{H}}_{G}, \rho \right] + \kappa \mathcal{L}[\hat{a}]\rho \, ,
\end{equation}
where the parametric Hamiltonian reads
\begin{align} \label{eq:parametric_oscillator_time}
\hat{\mathcal{H}}_{G} &= \omega_{c} \hat{a}^{\dagger}\hat{a} + \omega_{a} \hat{\sigma}_{+}\hat{\sigma}_{-} +  \lambda (\hat{a} \hat{\sigma}_{+} + \hat{a}^{\dagger} \hat{\sigma}_{-}) \nonumber \\ 
&+ \tfrac{1}{2}G( e^{2i\omega_{p}t}\hat{a}^{2} +  e^{-2i\omega_{p}t}\hat{a}^{\dagger 2}) \, ,
\end{align}
and the Lindblad superoperator $\mathcal{L}[\xi] \cdot = 2\xi\cdot\xi^{\dagger} -\cdot \xi^{\dagger}\xi - \xi^{\dagger}\xi\cdot$ accounts for losses in the form of photons leaving the subharmonic mode at a rate~$\kappa$. The explicit time-dependence of Eq.~(\ref{eq:parametric_oscillator_time}) is removed inside an interaction picture where the Hamiltonian becomes
\begin{equation} \label{eq:parametric_oscillator}
\hat{\mathcal{H}}_{G} = \Delta_{G} \hat{a}^{\dagger}\hat{a} + \Delta_{a} \hat{\sigma}_{+}\hat{\sigma}_{-} +  \lambda (\hat{a} \hat{\sigma}_{+} + \hat{a}^{\dagger} \hat{\sigma}_{-}) + \tfrac{1}{2}G(\hat{a}^{2} +  \hat{a}^{\dagger 2}) \, ,
\end{equation}
with detunings
\begin{align}
\Delta_{G} = \omega_{c}-\omega_{p} \, , \\
\Delta_{a}  = \omega_{a}-\omega_{p} \, .
\end{align}

%%%%%%%%%%%%%%%%%%%%%%%%%%%%%%%%%%%%%%%%%%%%%%%%%%%%%%%%%%%%%%%%%%%%%%%%%%%%%%%%%%%%%%%%%%%%%%%%%%%%%%
\subsection{Isomorphism between the JC--Rabi model and the qubit inside a parametric amplifier}
%%%%%%%%%%%%%%%%%%%%%%%%%%%%%%%%%%%%%%%%%%%%%%%%%%%%%%%%%%%%%%%%%%%%%%%%%%%%%%%%%%%%%%%%%%%%%%%%%%%%%%

The two-photon pump is responsible for a process of optical amplification that has been studied on detail in the absence of the qubit~\cite{Kimble_1986,Kimble_1987,Agarwal_libro}. Its effect on the subharmonic mode is to generate the squeezing transformation
$$\hat{S}(z) = \exp \tfrac{1}{2} (z^{*}\hat{a}^{2} - z\hat{a}^{\dagger 2})$$
with parameter
\begin{align}\label{eq:squeezing_parameter}
z = \frac14 \ln \left[\frac{\Delta_{G} - G}{\Delta_{G} + G} \right]  \, .
\end{align}

When a qubit is placed inside the parametric oscillator, it couples to the now squeezed mode. The qubit then probes the amplified quadratures of the field through rotating and counter-rotating terms equivalent to those found inside the JC-Rabi model. This allows the amplification process to connect the model Hamiltonians~(\ref{eq:Rabi_Hamiltonian}) and~(\ref{eq:parametric_oscillator}) through the unitary transformation
\begin{equation}\label{eq:squeezin_connection}
\mathcal{S}(z) \hat{\mathcal{H}}_{G} \mathcal{S}^{\dagger}(z) = \hat{\mathcal{H}}_{\eta} + \tfrac{1}{2} ( \Delta_{c} - \omega_{G})  \, ,
\end{equation}
with hyperbolic relations for the coupling strengths 
\begin{subequations}\label{eq:hyperbolic}
\begin{eqnarray}
\cosh z &=& \lambda^{\prime} / \lambda \, , \\
\sinh z &=& \eta \lambda^{\prime}  / \lambda \, ,
\end{eqnarray}
\end{subequations} 
and mode frequencies 
\begin{equation}\label{eq:frequency}
\omega_{c} = \sqrt{\Delta_{G}^{2} - G^{2}} \, .
\end{equation}

Equations (\ref{eq:squeezing_parameter})-(\ref{eq:frequency}) establish the isomorphism between the JC-Rabi model and a qubit inside a parametric cavity. They bring the ultra-strong coupling regime within reach, as mode and qubit frequencies are replaced by detunings to an external field~\cite{Leroux_2018}. Equation~(\ref{eq:squeezing_parameter}), in particular, also sets up a limiting pump amplitude $G_{\text{thr}} = \Delta_{G}$ where this isomorphism breaks down. The eigenstates of the squeezing operator are discrete and normalizable below this threshold and continuous and non-normalizable above it~\cite{Lo_1991}. While we have shown the isomorphism below threshold, the non-normalizable states represent an infinitely squeezed subharmonic mode whose photon intensity grows without bounds. Depletion of the pump is required to counteract this unphysical gain~\cite{Carmichael_2008}. This term breaks the connection between the parametric and the JC-Rabi models. Throughout this work we remain below threshold where both models are isomorphic and we are able to explore the JC-Rabi spectrum with $\eta \neq 1$. 

%%%%%%%%%%%%%%%%%%%%%%%%%%%%%%%%%%%%%%%%%%%%%%%%%%%%%%%%%%%%%%%%%%%%%%%%%%%%%%%%%%%%%%%%%%%%%%%%%%%%%%
\subsection{General properties of the eigenvalues and eigenstates of $\hat{\mathcal{H}}_{G}$ and $\hat{\mathcal{H}}_{\eta}$ }
%%%%%%%%%%%%%%%%%%%%%%%%%%%%%%%%%%%%%%%%%%%%%%%%%%%%%%%%%%%%%%%%%%%%%%%%%%%%%%%%%%%%%%%%%%%%%%%%%%%%%%

The eigenvalue problem according to Eq.~(\ref{eq:squeezin_connection}) is 
\begin{align}
\hat{\mathcal{H}}_{G} \vert \psi_{\beta} \rangle = E_{\beta}\vert \psi_{\beta} \rangle \, ,\\
\hat{\mathcal{H}}_{\eta} \vert \phi_{\beta} \rangle = \tilde{E}_{\beta} \vert \phi_{\beta}\rangle \, ,
\end{align}
where solutions to both models connect by
\begin{align}\label{eq:eigenvalue_problem}
&E_{\beta} = \tilde{E}_{\beta} + \tfrac{1}{2} ( \omega_{c} - \Delta_{G}) \, ,\\
&\vert \psi_{\beta} \rangle =  \mathcal{S}^{\dagger}(z) \vert \phi_{\beta} \rangle \, .
\end{align}

Since the parametric Hamiltonian commutes with the parity operator
\begin{equation}\label{eq:parity}
\hat{\Pi} = \exp[i\pi (\hat{a}^{\dagger}\hat{a} + \hat{\sigma}_{+}\hat{\sigma}_{-})] \, ,
\end{equation}
its eigenstates are classified into branches of even and odd parities according to the $\pm 1$ eigenvalues of $\hat{\Pi}$. All the bare states within each branch are coupled to different orders in the interaction due to the coexistence of coupling $\lambda$ and two-photon pump $G$ in the Hamiltonian. This leads to eigenstates of the form
\begin{subequations}
\begin{align}
\vert \psi_{\beta}^{\text{even}} \rangle &= \sum_{n=0}^{\infty} c_{\beta;g,2n} \vert g , 2n \rangle + c_{\beta;e,2n+1} \vert e , 2n+1 \rangle \, , \\
\vert \psi_{\beta}^{\text{odd}} \rangle &= \sum_{n=0}^{\infty} c_{\beta;g,2n+1} \vert g , 2n+1 \rangle + c_{\beta;e,2n} \vert e , 2n \rangle \, ,
\end{align}
\end{subequations}
with $\vert g \rangle$ and $\vert e \rangle$ the ground and excited states of the qubit, and $\vert n \rangle$ (with $n=0,1,\dots$) the photon number inside the mode. 

\begin{figure}[ht]
\begin{center}
\includegraphics[width=1.\linewidth]{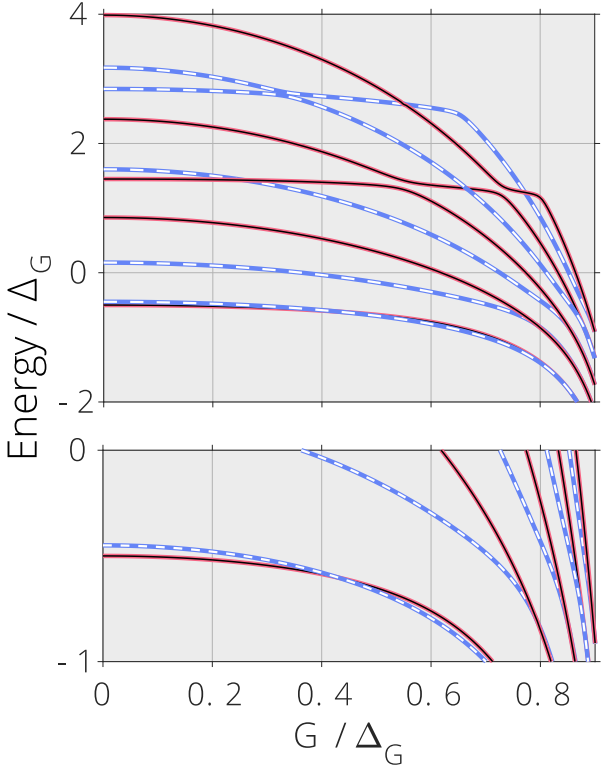}
\caption{Lowest eigenvalues of $\hat{\mathcal{H}}_{G}$ for $\lambda=0.95\Delta_{G}$ and $\Delta_{a}=\Delta_{G}$. Red solid (blue dashed) lines denote states of even (odd) parity.}\label{fig:eigenvalues}
\end{center}
\end{figure}

The coupling among bare states of equal parities is reflected on the eigenvalue spectrum, which displays avoided-crossings between states of the same parity and crossings between states of different parities. In Figure~\ref{fig:eigenvalues} we plot the lowest eigenvalues of $\hat{\mathcal{H}}_{G}$ as a function of the pump amplitude.  The results are obtained from numerical diagonalization using a truncated photon number basis for the parameters $\Delta_{a} = \Delta_{G}$ and $\lambda = 0.95\Delta_{G}$. This places us deep into the strong coupling regime where crossings and avoided-crossings of the spectrum are readily observed for small pump amplitudes. The first crossing is found at~\cite{Tomka_2014}
\begin{equation}
G = \frac{1}{\Delta_{a}}\sqrt{\Delta_{G}^{2}\Delta_{a}^{2}-\lambda^{4}}
\end{equation}
where the two lowest energy states meet.

\begin{figure}[ht]
\begin{center}
\includegraphics[width=1.\linewidth]{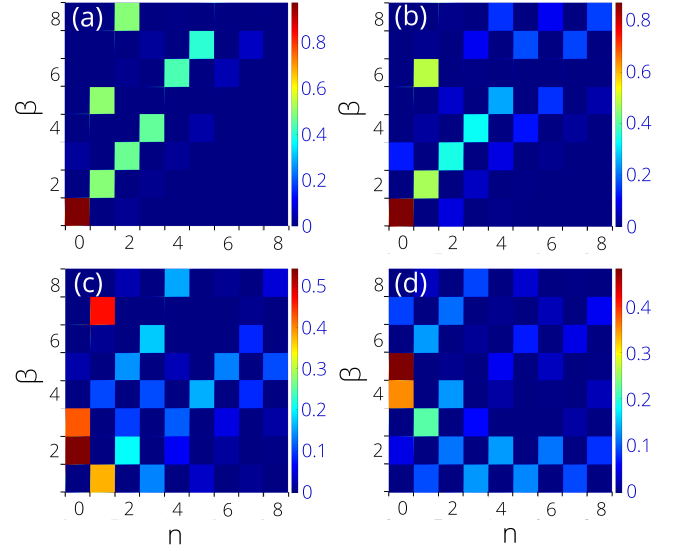}
\caption{Photon number population $\vert c_{\beta;g,n} \vert^{2}$ for the eigenstates of lowest energy of $\hat{\mathcal{H}}_{G}$ using the parameters of Fig.~\ref{fig:eigenvalues} with pump: $G/\Delta_{G} = 0.2$ (a), $0.4$ (b), $0.6$ (c), and $0.8$ (d). }\label{fig:eigenvectors}
\end{center}
\end{figure}

Changes in the spectrum with an increasing pump amplitude are attributed to a redistribution of the photon number population within each eigenstate. In Figure~\ref{fig:eigenvectors} we plot the populations $ \vert c_{\beta;g,n} \vert^{2}$ of the lowest-energy eigenstates for different pump amplitudes with $E_{\beta} < E_{\beta+1}$. For weak two-photon pump $(G \ll \lambda)$ the distributions tend to localize around a given photon number. This is shown in panel~(a) where the populations resemble those of the Jaynes--Cummings dressed states $(\vert n,\pm \rangle = \vert g,n\rangle \pm \vert e, n-1 \rangle)$. For example, states with $\beta = 2$ and $\beta = 5$ form the first JC doublet while $\beta = 3$ and $\beta = 8$ the second. As the pump is increased in panel~(b), the distributions broaden while keeping the same parity. Ultimately, the populations distribute among more and more photon numbers as shown in panels (c) and (d).

%%%%%%%%%%%%%%%%%%%%%%%%%%%%%%%%%%%%%%%%%%%%%%%%%%%%%%%%%%%%%%%%%%%%%%%%%%%%%%%%%%%%%%%%%%%%%%%%%%%%%%
\subsection{Dressed states at the first energy crossing}
%%%%%%%%%%%%%%%%%%%%%%%%%%%%%%%%%%%%%%%%%%%%%%%%%%%%%%%%%%%%%%%%%%%%%%%%%%%%%%%%%%%%%%%%%%%%%%%%%%%%%%

It is remarkable that the photon distribution can be obtained analytically at the first energy crossing~\cite{Braak_2011,Tomka_2014,Tomka_2015}. In the following we examine the eigenstates at this point using the isomorphism between the JC-Rabi and parametric models. 

Energy crossings signal a degeneracy in the system and have been shown to occur in the JC-Rabi model at the regular values~\cite{Tomka_2015}
\begin{equation}
\tilde{E}_{\text{cross}}(m) = m \omega_{c} - \frac{(1+\eta^{2})\lambda^{\prime 2}}{2\omega_{c}}   \, 
\end{equation} 
($m = 0 , 1, \dots$) under specific conditions for the parameters discussed on detail in Ref.~\cite{Tomka_2014}. The first crossing ($m=0$) is found under the condition
\begin{equation}\label{eq:cross_condition}
\omega_{a}  = (1 -\eta^{2})\lambda^{\prime 2}/\omega_{c} \, ,
\end{equation}
where the JC-Rabi Hamiltonian takes the simplified form
\begin{equation}\label{qpt_2}
\hat{\mathcal{H}}_{\eta} = \omega_{c} \hat{A}^{\dagger} \hat{A} - \frac{(1+\eta^{2})\lambda^{\prime 2}}{2\omega_{c}} \, .
\end{equation}
Here, $\hat{A}^{\dagger}$ and $\hat{A}$ are creation and annihilation operators represented by the matrices
\begin{align}\label{qpt_3}
\hat{A} = (\hat{A}^{\dagger})^{\dagger} = \frac{1}{\omega_{c}}\left( \begin{array}{cc}
\lambda^{\prime} & \omega_{c} \hat{a}  \\
\omega_{c} \hat{a} & \eta \lambda^{\prime}  \\
\end{array} \right) \, ,
\end{align}
that obey the commutation relation $[ \hat{A},\hat{A}^{\dagger}] = 1_{2\times 2}$ and account for the correlations that rise between subharmonic mode and qubit. 

The degenerate ground states of $\hat{\mathcal{H}}_{\eta}$ are readily obtained from the eigenvalue equation $$\hat{A} \vert \phi \rangle = 0$$ whose solutions lead to the dressed states
\begin{subequations}\label{eq:gatos}
\begin{align}
\vert \phi_{\text{even}}\rangle &= - \frac{ \vert g \rangle \vert C_{\alpha}^{+} \rangle - \sqrt{\eta} \vert e \rangle \vert C_{\alpha}^{-} \rangle}{\sqrt{\mathcal{N_{+}}}} \, , \label{eq:gatos_even}\\
\vert \phi_{\text{odd}} \rangle &= - \frac{ \vert g \rangle \vert C_{\alpha}^{-} \rangle - \sqrt{\eta} \vert e \rangle \vert C_{\alpha}^{+} \rangle}{\sqrt{\mathcal{N_{-}}}} \, , \label{eq:gatos_odd}
\end{align}
\end{subequations}
when diagonalized within the parity basis. States~(\ref{eq:gatos_even}) and (\ref{eq:gatos_odd}) describe a field in a positive or negative cat states correlated to the qubit state. Cat states can be written as
\begin{align}
\vert C_{\alpha}^{\pm} \rangle &= \left[ \hat{D}(\alpha) \pm \hat{D}(-\alpha) \right] \vert 0 \rangle \, 
\end{align}
where  $\hat{D}(\alpha) = \exp(-\alpha \hat{a}^{\dagger} + \alpha^{*} \hat{a} )$ is the displacement operator with amplitude
\begin{equation}\label{eq:amplitude}
\alpha = \sqrt{\eta} \lambda^{\prime} / \omega_{c} \, ,
\end{equation}
and the normalization $\mathcal{N}_{\pm}$ reflects the overlap between the coherent states forming each cat
\begin{equation}\label{eq:normalizacion}
\mathcal{N}_{\pm} = 2 [ (1 + \eta) \pm e^{-2 \vert \alpha \vert^{2}} (1-\eta) ] \, .
\end{equation}

It can already be seen that these dressed states play a central role in the optical response of a system satisfying the JC-Rabi model (see Sec.~\ref{sec:spectrum} below). The dressed states couple exclusively to one another under one-photon transitions 
\begin{subequations}\label{eq:channels}
\begin{align}
\hat{a} \vert \phi_{\text{even}}  \rangle = \alpha \sqrt{\frac{\mathcal{N}_{-}}{\mathcal{N}_{+}}} \vert \phi_{\text{odd}} \rangle \, , \\
\hat{a} \vert \phi_{\text{odd}}  \rangle = \alpha \sqrt{\frac{\mathcal{N}_{+}}{\mathcal{N}_{-}}}  \vert \phi_{\text{even}} \rangle \, .
\end{align}
\end{subequations} 
Thus forming a long--lived pair when we move into a dissipative setting. 

The eigenstates of $\hat{\mathcal{H}}_{G}$ can also be obtained at the crossing using Eqs.~(\ref{eq:eigenvalue_problem}) and~(\ref{eq:gatos}). The eigenstates take the form
\begin{subequations}\label{eq:gatos_sq}
\begin{align}
\vert \psi_{\text{even}} \rangle &= - \frac{\vert g \rangle \vert \tilde{C}^{+}_{\alpha , z} \rangle - \sqrt{\eta} \vert e \rangle \vert \tilde{C}^{-}_{\alpha , z} \rangle}{\sqrt{\mathcal{N_{+}}}} \, , \label{eq:gatos_even_sq}\\
\vert \psi_{\text{odd}} \rangle &= -  \frac{\vert g \rangle \vert \tilde{C}^{-}_{\alpha , z} \rangle - \sqrt{\eta} \vert e \rangle \vert \tilde{C}^{+}_{\alpha , z} \rangle}{\sqrt{\mathcal{N_{-}}}} \, . \label{eq:gatos_odd_sq}
\end{align}
\end{subequations}
where the field is described by a superposition of two-photon coherent states~\cite{Yuen_1976,Mandel_1995}
\begin{equation}
\vert \tilde{C}^{\pm}_{\alpha , z} \rangle = \left[S^{\dagger}(z) \hat{D}(\alpha) \pm S^{\dagger}(z) \hat{D}(-\alpha) \right] \vert 0 \rangle \, .
\end{equation}
Due to the effect of the squeezing operator, these states couple to states outside the pair under one-photon transitions. The probability amplitude to remain inside the pair after is given by
\begin{subequations}\label{eq:channels_sq}
\begin{align}
&\frac{\langle {\psi}_{\text{odd}} \vert \hat{a} \vert {\psi}_{\text{even}}  \rangle}{\sqrt{\langle {\psi}_{\text{even}} \vert \hat{a}^{\dagger}\hat{a} \vert {\psi}_{\text{even}  }\rangle}} =  \frac{\mathcal{N}_{-} \alpha \cosh z - \mathcal{N}_{+} \alpha^{*} \sinh z }{\sqrt{\mathcal{N}_{-} O_{+}}}  \,  \\
&\frac{\langle {\psi}_{\text{even}} \vert \hat{a} \vert {\psi}_{\text{odd}}  \rangle}{\sqrt{\langle {\psi}_{\text{odd}} \vert \hat{a}^{\dagger}\hat{a} \vert {\psi}_{\text{odd}  }\rangle}} =  \frac{\mathcal{N}_{+} \alpha \cosh z - \mathcal{N}_{-} \alpha^{*} \sinh z }{\sqrt{\mathcal{N}_{+} O_{-}}}  \,  
\end{align}
\end{subequations} 
where
\begin{align}
O_{\pm} =  \mathcal{N}_{\pm} \sinh^{2}z &+ \tfrac{1}{2}(\mathcal{N}_{+} + \mathcal{N}_{-})\vert \alpha \cosh z  - \alpha^{*} \sinh z \vert^{2} \nonumber \\
&\mp \tfrac{1}{2}(\mathcal{N}_{+} - \mathcal{N}_{-})\vert \alpha \cosh z  + \alpha^{*} \sinh z \vert^{2} \nonumber
\end{align}
is proportional to the photon number expectation within each state.

%%%%%%%%%%%%%%%%%%%%%%%%%%%%%%%%%%%%%%%%%%%%%%%%%%%%%%%%%%%%%%%%%%%%%%%%%%%%%%%%%%%%%%%%%%%%%%%%%%%%%%
\section{Probed spectrum of the JC-Rabi model via the parametric Hamiltonian}\label{sec:spectrum}
%%%%%%%%%%%%%%%%%%%%%%%%%%%%%%%%%%%%%%%%%%%%%%%%%%%%%%%%%%%%%%%%%%%%%%%%%%%%%%%%%%%%%%%%%%%%%%%%%%%%%%

We now move on to simulate the energy spectrum of Figure~\ref{fig:eigenvalues} as probed by a coherent beam with tunable frequency exciting the subharmonic mode. The probe has an amplitude $\epsilon$ and is detuned a frequency $\delta$ from the pump beam. Its effect over the system dynamics is given through an additional term  
\begin{equation}
\hat{\mathcal{H}}_{\text{probe}} = \epsilon \left( \hat{a} e^{i \delta t} + \hat{a}^{\dagger} e^{-i \delta t} \right) \, ,
\end{equation}
inside the master equation~(\ref{eq:master_eq}). 
 
For weak amplitudes $(\epsilon \ll G,\lambda)$ the probe creates one-photon channels that couple two eigenstates $\vert \psi_{\beta} \rangle$ and $\vert \psi_{\beta^{\prime}} \rangle$  of opposite parities separated by an energy difference $\Delta E= E_{\beta}-E_{\beta^{\prime}}$. The maximum transition probability is reached when the resonance condition $$ \delta = \Delta E  $$ is met and is weighted by the transition matrix element $\langle \psi_{\beta^{\prime}} \vert \hat{a} \vert \psi_{\beta} \rangle$ between the states. Figures~\ref{fig:eigenvalues} and~~\ref{fig:eigenvectors} provide a reference for the energy separation and the matrix element, respectively. 

\begin{figure}[ht]
\begin{center}
\includegraphics[width=1.\linewidth]{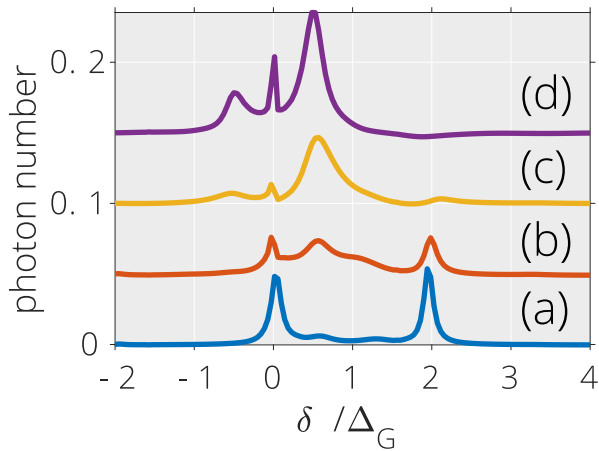}
\caption{Resonance spectrum for $\Delta_{a}=\Delta_{G}$, $\lambda=0.95\Delta_{G}$ and $\kappa = 0.1\Delta_{G}$ probed by a field of amplitude $\epsilon = 0.03\Delta_{G}$. The lines correspond to the same pump amplitudes as Figure~\ref{fig:eigenvectors}. Results are displaced vertically to account for the background photon number and allow for clarity of presentation}\label{fig:spectrum}
\end{center}
\end{figure}

In Figure~\ref{fig:spectrum} we plot the steady-state photon number expectation obtained from a numerical evolution of the master equation~(\ref{eq:master_eq}) with initial state $\vert g ,0 \rangle$. The results are shown for the (a)-(d) lines of Figure~\ref{fig:eigenvalues} as probed by a weak coherent field ($\epsilon \simeq 0.03 \lambda$) inside the strong coupling regime ($\kappa \simeq \lambda/10$). For each line we find several peaks that are traced back to transitions between eigenstates $\beta$ and $\beta^{\prime}$ of $\hat{\mathcal{H}}_{G}$. We begin with line (a) where the $\beta = 1$ to $\beta^{\prime}=2$ and $\beta = 1$ to $\beta^{\prime}=5$ transitions are resolved by two peaks separated a distance~$\simeq2\lambda$. These transitions correspond to the JC doublet made available by the weak pump amplitude~$G$. We consider next line (b) where an additional peak at $\delta \simeq 0.7 \Delta_{G}$ is resolved. The peak corresponds to the $\beta = 2$ to $\beta^{\prime} =3$ transition and its appearance is attributed to a broadening of the photon distributions within these two eigenstates. The matrix element connecting two states increases when the distributions broaden. In line (c) the pump is further ramped up, leading to a peak around $\delta \simeq -0.5 \Delta_{G}$ and a broadened line around $\delta \simeq 0.5 \Delta_{G} $. The peak at negative frequency corresponds to the $\beta = 3$ to $\beta^{\prime} = 1$ transition and its sign follows from the large overlap between $\vert \psi_{\beta=3}\rangle$ and the initial state $\vert g ,0 \rangle$. The broadened peak appears as several transitions merge, \textit{e.g.}, the $\beta = 2$ to $\beta=4 $ $(\delta \simeq 0.7 \Delta_{G})$ and $\beta = 3$ to $\beta=4$ $(\delta \simeq 0.25 \Delta_{G})$. Finally, in line (d) we can see a combination of all these effects.

\begin{figure}[ht]
\begin{center}
\includegraphics[width=1.\linewidth]{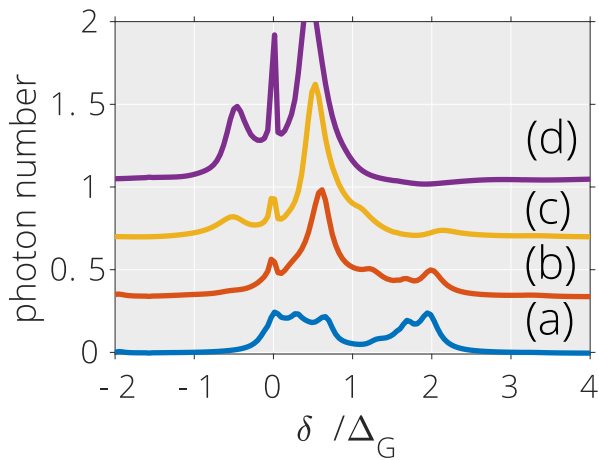}
\caption{Resonance spectrum for the same system parameters as Fig.~\ref{fig:spectrum} with probing field $\epsilon = 0.125\Delta_{G}$. The maximum photon expectation reached in (d) is $\simeq 2.165$. }\label{fig:spectrum_multi}
\end{center}
\end{figure}

For stronger probe amplitudes $\epsilon$ it is possible to connect states of the same parity through $2m$-photon transitions~\cite{Agarwal_1991,Shamailov_2010,Rempe_2017}. The $m$-th photon transition probability is maximized for the resonance condition
$$ m \delta = \Delta E .$$
In Figure~\ref{fig:spectrum_multi} we plot the absorption spectrum probed by a driving field of ampliude $\epsilon \simeq 0.12\lambda$ for the same parameters as Figure~\ref{fig:spectrum}. We begin with line (a) again, where additional peaks near frequencies $\delta/\Delta_{G} \simeq 0.3 $, $1.7$ are resolved. These peaks correspond to two-photon transitions between states of the same parity ($\beta = 1$ to $\beta^{\prime}= 3$ and $\beta=1$ to $\beta^{\prime}= 8$). Notice also a peak at $\delta \simeq 0.7 \Delta_{G}$ corresponding to the one-photon transition $\beta = 2$ to $\beta^{\prime} = 3$. Two-photon transitions in line (b) can still be resolved, as the peak around $\delta \simeq 1.7 \Delta_{G}$ corresponds to the transition $\beta = 1$ to $\beta^{\prime}=9$. Since one-photon transitions are power broadened by the intense probe we are unable to resolve other peaks. This is exemplified in lines (c) and (d) where one-photon transitions dominate and lines resembling those of Figure~\ref{fig:spectrum} are found, albeit with increased amplitude due to a stronger probe beam. 

%%%%%%%%%%%%%%%%%%%%%%%%%%%%%%%%%%%%%%%%%%%%%%%%%%
\section{Ground states outside the energy crossing: Role of dissipative processes}\label{sec:outside}
%%%%%%%%%%%%%%%%%%%%%%%%%%%%%%%%%%%%%%%%%%%%%%%%%%

We have shown that the JC-Rabi model is isomorphic to a model Hamiltonian describing a qubit inside a parametric cavity, a system that can be realized under current experimental constrains. We have also simulated the absorption spectra of this realizable system and connected the different resonance peaks to transitions between eigenstates of the model Hamiltonian. In doing so we have shown how to probe the spectrum of the JC-Rabi model. The next section is concerned with the narrow peak found around $\delta =0$ in the absorption spectra of Figures~\ref{fig:spectrum} and~\ref{fig:spectrum_multi}. In particular, we are interested in the eigenstates that lead to this peak. Over the next section we show that these low-energy states resemble the dressed states of Eqs.~(\ref{eq:gatos}) and~(\ref{eq:gatos_sq})---even though the system is far away from the first crossing. For this, we construct an ansatz $\vert \phi_{ans} \rangle$ based on the dressed states and find that the system overlaps significantly with this ansatz as it evolves in time.

The evolution is taken using quantum trajectories equivalent to the master equation~(\ref{eq:master_eq}). In quantum trajectory theory the density matrix $\rho$ is replaced by an ensemble of stochastic wave functions $\vert \psi_{c} (t) \rangle$ conditioned to a particular measurement record. The wave functions evolve under the Schr\"{o}dinger equation
\begin{equation}\label{eq:trajectory}
i\hbar \vert \dot{\psi}_{c}(t) \rangle = (\hat{\mathcal{H}}_{G} -i\kappa \hat{a}^{\dagger}\hat{a}) \vert {\psi}_{c}(t) \rangle
\end{equation}
and are interrupted by the action of the jump operator $\hat{a}$ when a photo electron is detected at times determined in a Monte Carlo fashion~\cite{Carmichael_2008}. This allows us to see the changes in the parity of the system each time a photo electron is detected. 

The results for the parametric model are to be compared to those of the JC-Rabi model. This procedure brings to light an important aspect of the current approach: the surrounding environment is coupled to a particular set of modes. In the parametric model the environment is coupled to the subharmonic mode of the cavity and the evolution is described by
\begin{equation}\label{eq:master_eq_2}
\dot{\rho} = -i \left[ \hat{\mathcal{H}}_{G}, \rho \right] + \kappa \mathcal{L}[\hat{a}]\rho \, .
\end{equation}
By comparison, in the JC-Rabi Hamiltonian generated through coupled lambda transitions the master equation reads
\begin{equation}\label{eq:master_eq_1}
\dot{\rho} = -i \left[ \hat{\mathcal{H}}_{\eta}, \rho \right] + \kappa \mathcal{L}[\hat{a}]\rho \, ,
\end{equation}
and the environment is coupled to the cavity mode. While $\hat{\mathcal{H}}_{G}$ and $\hat{\mathcal{H}}_{\eta}$ are isomorphic, the master equations~(\ref{eq:master_eq_2}) and~(\ref{eq:master_eq_1}) are not, there is a two-photon decay missing.

%%%%%%%%%%%%%%%%%%%%%%%%%%%%%%%%%%%%%%%%%%%%%%%%%%
\subsection{Ground states of the coupled lambda transitions}
%%%%%%%%%%%%%%%%%%%%%%%%%%%%%%%%%%%%%%%%%%%%%%%%%%

We begin with the JC-Rabi model where the system evolves under Eq.~(\ref{eq:master_eq_1}) and propose the ansatz 
%$$\vert \phi_{\text{ans}} \rangle = \frac{\vert \phi_{\text{even}}\rangle + \vert \phi_{\text{odd}}\rangle}{\sqrt{2}}$$ with
\begin{subequations}\label{eq:gatos_ans}
\begin{align}
\vert \phi_{\text{even}}\rangle &= - \frac{ \vert g \rangle \vert C_{\alpha_{\text{ans}}}^{+} \rangle - \sqrt{\eta} \vert e \rangle \vert C_{\alpha_{\text{ans}}}^{-} \rangle}{\sqrt{\mathcal{N_{+}}}} \, , \label{eq:gatos_even_ans}\\
\vert \phi_{\text{odd}} \rangle &= - \frac{ \vert g \rangle \vert C_{\alpha_{\text{ans}}}^{-} \rangle - \sqrt{\eta} \vert e \rangle \vert C_{\alpha_{\text{ans}}}^{+} \rangle}{\sqrt{\mathcal{N_{-}}}} \, . \label{eq:gatos_odd_ans}
\end{align}
\end{subequations}
These states have the same form as Eq.~(\ref{eq:gatos}), but with corrected field amplitudes
\begin{equation}\label{eq:amplitude_ans}
\alpha_{\text{ans}} = \sqrt{\eta} \lambda^{\prime} / (\omega_{c}  + i \kappa)
\end{equation}
to account for the decay rate and displacements from the crossing. The ansatz is built from the stationary value of a driven-damped harmonic oscillator and our observations of different quantum trajectories. For convenience we define the projector
\begin{equation}\label{eq:prob}
\hat{\mathcal{P}}_{\text{a}} = \vert \phi_{\text{even}}\rangle \langle \phi_{\text{even}}\vert +   \vert \phi_{\text{odd}}\rangle \langle \phi_{\text{odd}}\vert \, ,
\end{equation}
such that $P_{\text{a}}=\text{Tr}[\hat{\mathcal{P}}_{\text{a}}\vert \phi_{c} \rangle \langle \phi_{c} \vert]$ gives the probability to find the system inside the ansatz.

\begin{figure}[ht]
\begin{center}
\includegraphics[width=.9\linewidth]{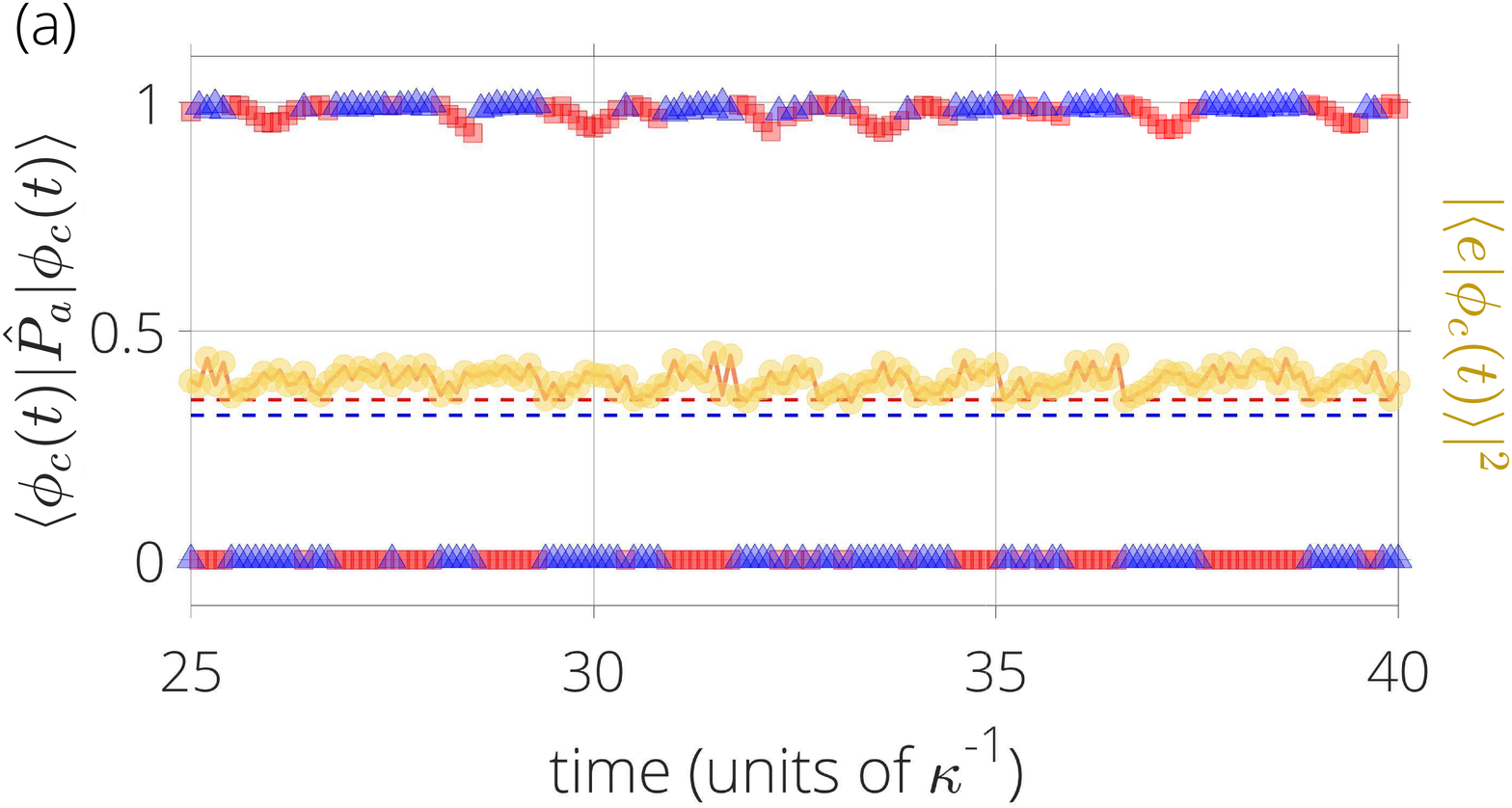} \\
\includegraphics[width=1.\linewidth]{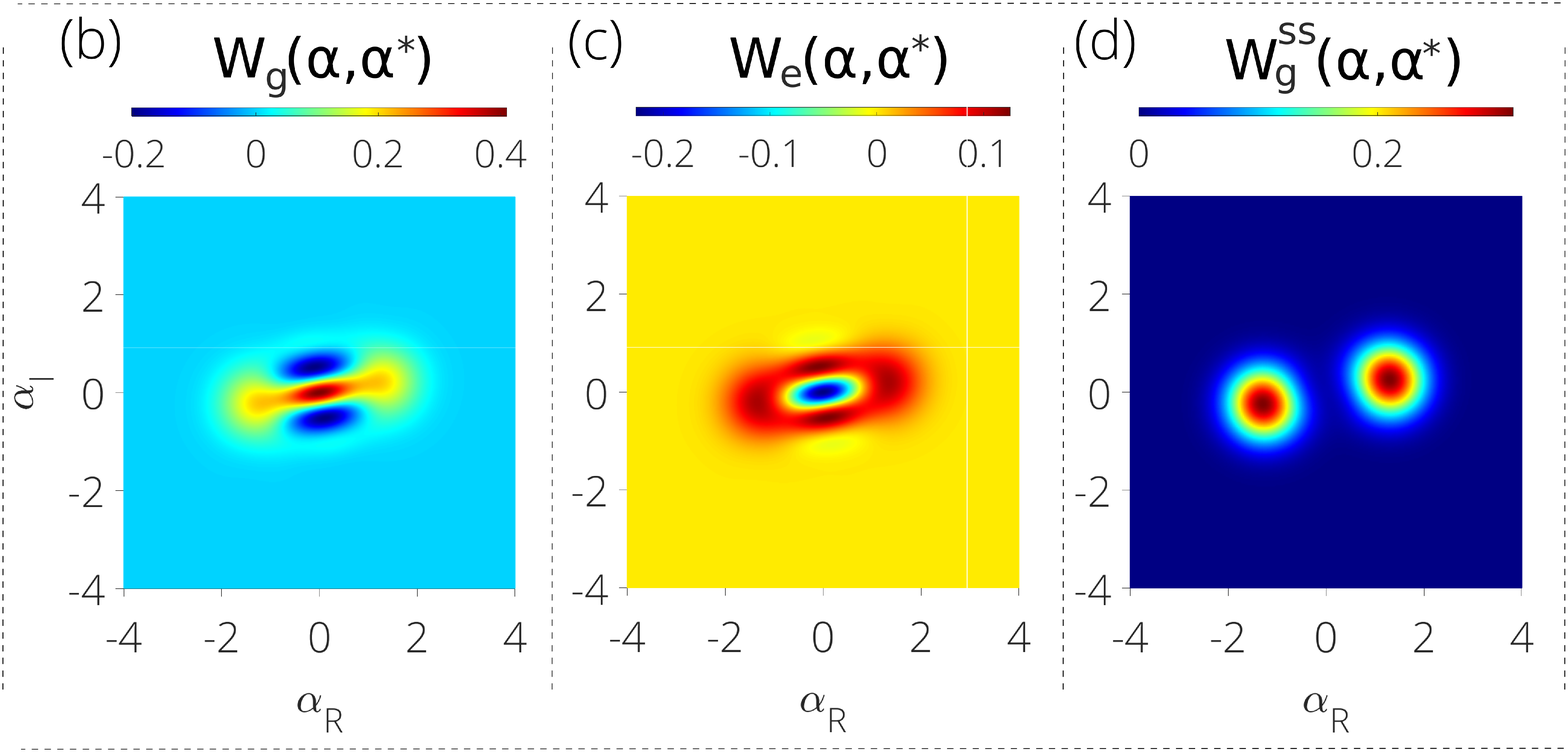}
\caption{Validity of the state ansatz~(\ref{eq:gatos_ans}) for $\lambda = 0.95\Delta_{G}$, $\Delta_{G} = \Delta_{a}$, $\kappa = 0.1 \Delta_{G}$, and $G=0.8\Delta_{G}$ (equivalent to $\eta\simeq 0.5$). (a) Probabilities $\vert \langle \phi_{\text{even}} \vert \phi_{c} \rangle \vert^{2} $ (red squares), $\vert \langle \phi_{\text{odd}} \vert \phi_{c} \rangle \vert^{2}$ (blue triangles) and $\vert \langle e \vert \phi_{c} \rangle \vert^{2}$ (yellow circles). (b)-(c) Wigner distributions conditioned to the ground and excited state of the qubit following a jump to a state of odd parity. (d) Wigner distribution obtained from the steady state of master equation~(\ref{eq:master_eq_1}).}\label{fig:traj_1}
\end{center}
\end{figure}

The probability $P_{\text{a}}$ is plotted in Figure~\ref{fig:traj_1}a for a sample quantum trajectory with initial state $\vert \phi_{c}(t_{o})\rangle = \vert g , 0 \rangle$ using the parameters that lead to line (d) above. The probability is decomposed into red squares and blue triangles that denote, respectively, the terms $\vert \langle \phi_{\text{even}} \vert \phi_{c} \rangle \vert^{2} $ and $\vert \langle \phi_{\text{odd}} \vert \phi_{c} \rangle \vert^{2}$ of Eq.~(\ref{eq:prob}). Notice that the conditioned wave function settles near the ansatz states and changes its parity each time a photo electron is detected. The wave function also oscillates between two detection events with a larger amplitude on the even branch. This oscillation is caused by overlap between the wave function and excited states outside the low-energy pair. Excited states of even and odd parity overlap significantly with $\vert g ,0 \rangle$ ($\vert e ,1 \rangle$) and $\vert e ,0 \rangle$ ($\vert g ,1 \rangle$) for this set of parameters due to the low photon number expectation ($\langle \hat{a}^{\dagger}\hat{a}\rangle_{\text{ss}} \simeq 1.42$). Such that the wave function is sent into a superposition of ansatz and excited states after the detection of a photo electron that explains the oscillations seen in the figure. 
%An additional source of error can rise from the ansatz itself. The ansatz is built from the stationary value of a driven harmonic oscillator and, as such, can not be expected to capture the transient dynamics that accompany the evolution of the system. During intervals of no-photon detection the amplitude of a coherent state is reduced to $e^{-\kappa t}\alpha $ a consequence of the information gathered from the system during these intervals. 
%

These effects are reflected on the state of the qubit and mode. The probability to find the qubit on the excited state is displayed by yellow circles on Fig.~\ref{fig:traj_1} where it is seen to fluctuate around two separate values. The large fluctuations denote a change on the qubit as the wave function jumps between even and odd subspaces, and signal the overlap between the two coherent states forming each cat state [see $\mathcal{N}_{\pm}$ in Eq.~(\ref{eq:gatos_ans})]. This overlap can be visualized using the Wigner distribution of the field $$W_{l}(\alpha, \alpha^{*}) = \frac{1}{\pi^{2}}\text{Tr}\left[ \rho_{l} \int d^{2}\xi e^{\xi^{*}(\alpha - \hat{a})- \xi(\alpha^{*} - \hat{a}^{\dagger})} \right] ,$$ with $l= \lbrace g,e \rbrace$ and $\rho_{l} = \text{Tr}_{\text{q}}[\vert \phi_{c} \rangle \langle \phi_{c} \vert l \rangle \langle l \vert ]$ the density matrix of the field conditioned to the excited or ground states of the qubit. The Wigner distributions are plotted on Figs.~\ref{fig:traj_1}b and~\ref{fig:traj_1}c after the system is driven to the odd subspace by the detection of a photo electron. In this example $W_{g}$ is found inside a negative cat state while $W_{e}$ inside a positive cat state. Each time the system changes parity the conditioned distributions switch. By performing a time average over many of these changes the interference fringes dissapear and give way to the statistical mixture shown in Fig.~\ref{fig:traj_1}d. 

\begin{figure}[ht]
\begin{center}
\includegraphics[width=.9\linewidth]{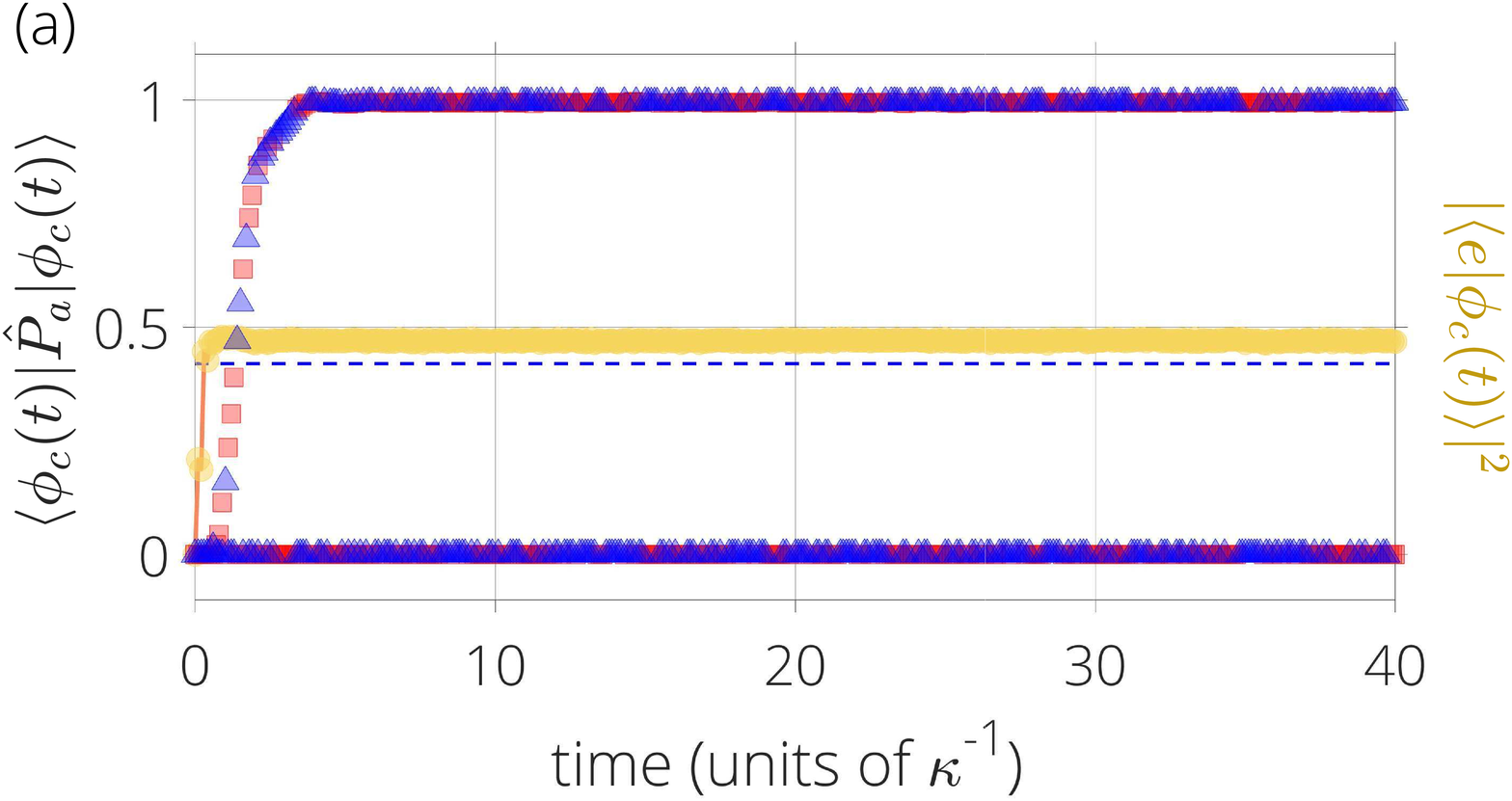} \\
\includegraphics[width=.9\linewidth]{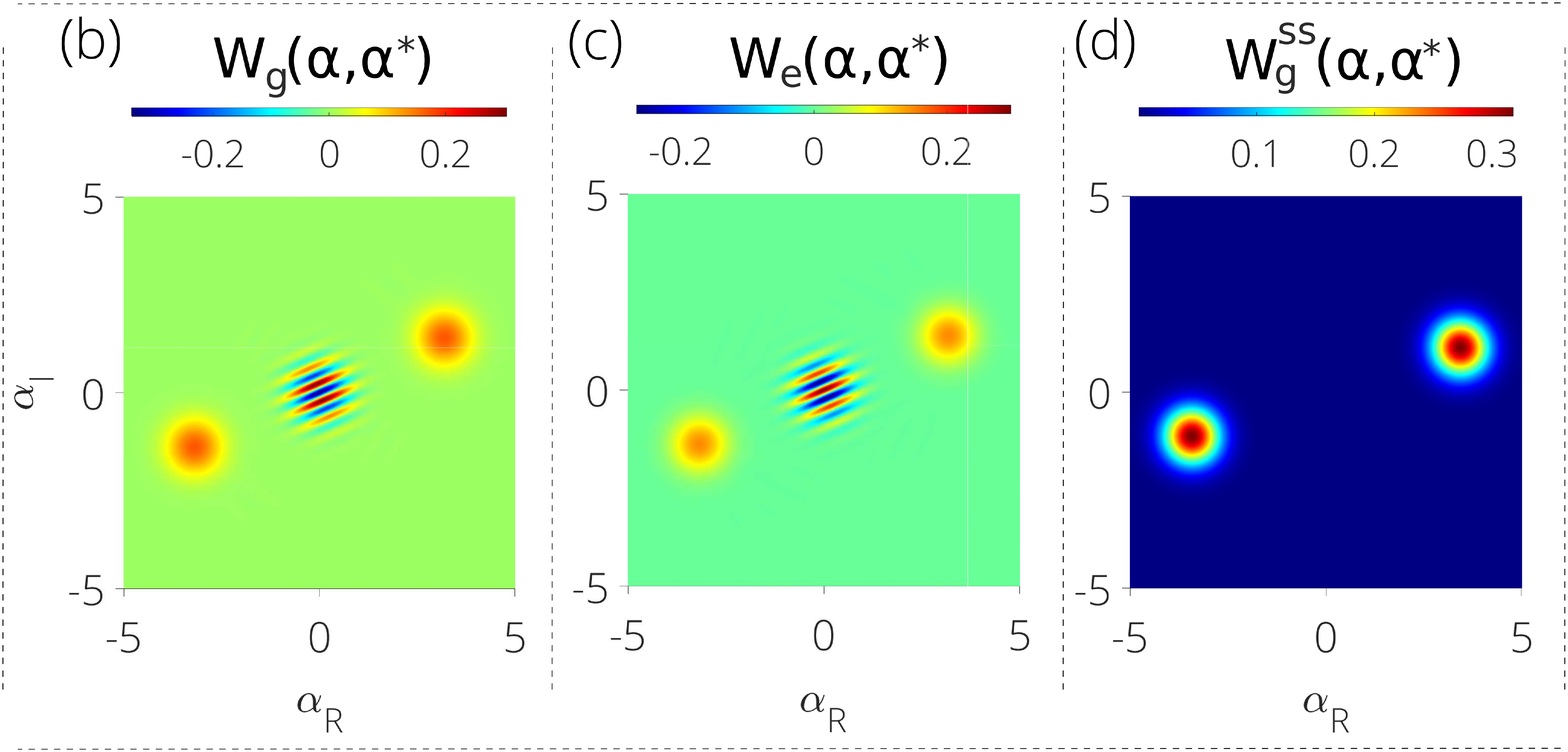}
\caption{Validity of the state ansatz~(\ref{eq:gatos_ans}) for $\lambda = 0.95\Delta_{G}$, $\Delta_{G} = \Delta_{a}$, $\kappa = 0.1 \Delta_{G}$,  and $G=0.95\Delta_{G}$ (equivalent to $\eta\simeq 0.72$). (a) Probabilities $\vert \langle \phi_{\text{even}} \vert \phi_{c} \rangle \vert^{2} $ (red squares), $\vert \langle \phi_{\text{odd}} \vert \phi_{c} \rangle \vert^{2}$ (blue triangles) and $\vert \langle e \vert \phi_{c} \rangle \vert^{2}$ (yellow circles). (b)-(c) Wigner distributions following a jump to a state of even parity conditioned to the ground and excited state of the qubit. (d) Wigner distribution obtained from the steady state of master equation~(\ref{eq:master_eq_1}). }\label{fig:many_photons}
\end{center}
\end{figure}

The ansatz remains valid as we move further away from the energy crossing. In Figure~\ref{fig:many_photons} we illustrate the case of $\eta \simeq 0.72$ where the steady-state displays a higher photon-number expectation. The amplitude of the oscillations outside the ansatz is reduced in this case, as well as the fluctuations of the qubit state. Since the probability of finding the systems in either dressed state is not unity fluctuations can still drive the system outside the pair for relatively short times.

%%%%%%%%%%%%%%%%%%%%%%%%%%%%%%%%%%%%%%%%%%%%%%%%%%
\subsection{Qubit inside a parametric amplifier}
%%%%%%%%%%%%%%%%%%%%%%%%%%%%%%%%%%%%%%%%%%%%%%%%%%

The previous results explain the narrow resonance found in Figs.~\ref{fig:spectrum} and~\ref{fig:spectrum_multi}. The conditioned wave function of the system is composed predominantly of two dressed states that couple to one another under one-photon transitions. This creates a long lived pair that displays a high degree of coherence and manifests as a narrow resonance. We now extend the analysis for the qubit inside a parametric amplifier using equation~(\ref{eq:master_eq_2}). In this case the conditioned wave function is composed  predominantly of the states
\begin{subequations}
\begin{align}
\vert \psi_{\text{even}} \rangle &= - \frac{\vert g \rangle \vert \tilde{C}^{+}_{\alpha , z} \rangle - \sqrt{\eta} \vert e \rangle \vert \tilde{C}^{-}_{\alpha , z} \rangle}{\sqrt{\mathcal{N_{+}}}} \, , \nonumber \\
\vert \psi_{\text{odd}} \rangle &= -  \frac{\vert g \rangle \vert \tilde{C}^{-}_{\alpha , z} \rangle - \sqrt{\eta} \vert e \rangle \vert \tilde{C}^{+}_{\alpha , z} \rangle}{\sqrt{\mathcal{N_{-}}}} \, . \nonumber 
\end{align}
\end{subequations}
We consider a qualitative description of the conditioned wave function in the following, as two-photon coherent states $\vert \tilde{C}^{\pm}_{\alpha , z} \rangle$ display changes in both squeezing factor $z$ and amplitude $\alpha$ when evolving under a quadratic Hamiltonians~\cite{Yuen_1976}. With an interplay between these two free parameters and the uncertainty relation
\begin{equation}
\langle 0 \vert \hat{D}^{\dagger}(\alpha) S(z)  (\Delta\hat{a})^{2}  S^{\dagger}(z) \hat{D}(\alpha) \vert 0 \rangle = -\cosh z \sinh z 
\end{equation}
it is challenging to determine the parameters for the corrected field equivalent to Eq.~(\ref{eq:amplitude_ans}). 

\begin{figure}
\includegraphics[width=.9\linewidth]{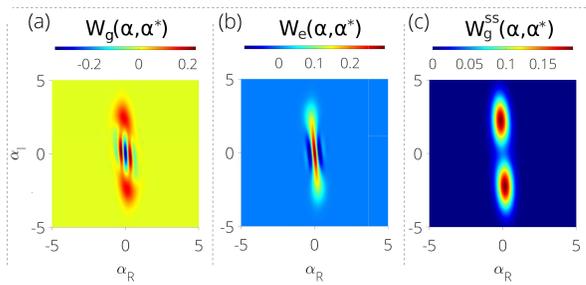}
\caption{(a)-(b) Wigner distributions following a jump to a state of even parity conditioned to the ground and excited state of the qubit. (c) Wigner distribution obtained from the steady state of master equation~(\ref{eq:master_eq_2}).}\label{fig:Wigner_comparison}
\end{figure}

In Figure~\ref{fig:Wigner_comparison} we show the conditioned Wigner distributions of the field after it is driven to a state of even parity by detection of a photo electron. The distributions are conditioned to the ground and excited state of the qubit and show two-photon coherent states. While the parameters $z$ and $\alpha$ are not characterized---they fluctuate as the wave function evolves---the two distributions remain correlated at all times. The fact that both distributions evolve as a correlated pair of two-photon states means that they couple predominantly to one another under Eq.~(\ref{eq:channels_sq}) and explain the narrow peak.

%%%%%%%%%%%%%%%%%%%%%%%%%%%%%%%%%%%%%%%%%%%%%
\section{Discussion}\label{sec:discussion}
%%%%%%%%%%%%%%%%%%%%%%%%%%%%%%%%%%%%%%%%%%%%%

%This manuscript treats the JC-Rabi spectrum and how it can be probed using an isomorphic model. 
By probing the spectrum of the qubit inside a parametric amplifier---and by consequence that of the JC--Rabi model---we have opened a window into the competing processes that rule the dynamics for each model. This competition occurs between: (i) coupling strength $\lambda$ and two-photon pump $G$ for the parametric model; and (ii) rotating terms $\lambda^{\prime}$ and counter-rotating terms $\eta \lambda^{\prime}$ for the JC-Rabi model. The competition is apparent under a dissipative setting where the system settles into a steady state that minimizes fluctuations related to the dominating process and, for this system, manifests as a narrow resonance line.

Take the parametric model as an example. In the absence of a pump ($G=0$) the eigenstates of $\hat{\mathcal{H}}_{G}$ are the Jaynes--Cummings dressed states $\vert n,\pm \rangle = \vert g,n\rangle \pm \vert e, n-1 \rangle$. These eigenstates are organized into manifolds with a given excitation number $\hat{N} = \hat{a}^{\dagger}\hat{a} + \hat{\sigma}_{+} \hat{\sigma}_{-} $ such that, when dissipation is considered, the system settles into the absolute ground state at long times. By comparison, in the absence of the qubit ($\lambda = 0$) the eigenstates of $\hat{\mathcal{H}}_{G}$ are squeezed states. It has been discussed by Carmichael~\cite{Carmichael_2008b} that when dissipation is taken into account the steady-state of a parametric cavity with adiabatic elimination of the pump takes the approximate form
\begin{equation}
\rho^{ss} = \tfrac{1}{2} \left[ p_{\text{even}} \vert C_{\alpha}^{+} \rangle \langle C_{\alpha}^{+} \vert + p_{\text{odd}} \vert C_{\alpha}^{-} \rangle \langle C_{\alpha}^{-} \vert \right]
\end{equation} 
in the limit of small system size when two-photon decay dominates over single photon decay. Cat states $\vert C_{\alpha}^{\pm} \rangle$ are ubiquitous in systems driven by two-photon processes, as they are eigenstates of the parity $e^{i\hat{n}}$ and two-photon annihilation $\hat{a}^{2}$ operators~\cite{Agarwal_1993,Gilles_1998,Ashhab_2010,Minganti_2016,Maldonado_2019}. By settling into this pair, the system minimizes fluctuations of the two-photon operator. When both pump and qubit are considered, the steady state settles into a dressed state pair displaying a combination of these two effects [see Eqs.~(\ref{eq:gatos_even_sq}) and~(\ref{eq:gatos_odd_sq})]. The combination was already apparent in Fig.~\ref{fig:eigenvectors} where the states transitioned from having a well-defined excitation number to having a broad photon number distribution as the pump was ramped up. It is worth noticing that these general features appear in the parametrically driven Kerr-oscillator~\cite{Minganti_2016} where the effect of the qubit is replaced by a nonlinear material.

The role of fluctuations in defining the steady state helps to explain the structure of the lowest energy states of both the Jaynes--Cummings-Rabi model and the parametric model. In particular, it helps to explain why the analytical structure found at the energy crossing is mantained far away from this point. Dressed states where the field is inside a cat state conditioned to the state of the qubit are the natural way to combine the two-excitation processes introduced by an external drive and coupling between qubit and field. In addition, the simplicity of the ansatz for the JC--Rabi model (and the large overlap it displays with the ground state) suggest that an analytical result can be obtained using a variational approach with the amplitude $\alpha$ as a variational parameter. This is left for future work. 

%%%%%%%%%%%%%%%%%%%%%%%%%%%%%%%%%%%%%%%%%
\section{Conclusion}\label{sec:conclusion}
%%%%%%%%%%%%%%%%%%%%%%%%%%%%%%%%%%%%%%%%%

We have presented an isomorphism between a qubit inside a parametric amplifier and the Jaynes-Cummings-Rabi model. The isomorphism brings the ultra-strong coupling regime within reach, as mode and qubit frequencies are replaced by detunings to an external field and coupling rates are managed by the amplitude of this field. We simulated spectrum of this system as probed by a coherent beam with a varying frequency. The peaks of the absorption spectrum where then connected to different transitions among eigenstates of the model; thus showing how the JC-Rabi spectrum can be probed under current experimental settings. 

Deep into the strong-coupling regime the spectrum displays a narrow peak whose breath decreases as more photons are injected to the subharmonic mode. The peak is attributed to a transition between two low energy states of the system. In particular, we have shown that these states form a long-lived pair under one-photon transitions as those induced by the probe and environment. The structure of this pair is kept accross the parameter space and provides a window into the underlying processes that rule the JC-Rabi model.

\section{Acknowledgements}

GSA thanks the AFOSR award no FA9550-20-1-0366 for supporting this work. R.G.-J. acknowledges financial support by the National Science Foundation QII-TAQS (Award No. 1936359).
%%%%%%%%%%%%%%%%%%%%%%%%%%%%%%%%%%%%


\begin{thebibliography}{1}
%%%%%%%%%%%%%%%%%%%%%%%%%%%%%%%%%%%%


\bibitem{Cohen_1998} C.~Cohen-Tannoudji, J.~Dupont-Roc, and G.~ Grynberg: \textit{Atom-photon interactions : basic processes and applications} (Wiley, New York, 1998).
% Bloch
\bibitem{Bloch_1940} F.~Bloch and A.~Siegert \textit{Phys. Rev.} \textbf{57}, 522 (1940).

% Bloch Siegert shift
\bibitem{Shirley_1965} J.~H.~Shirley, \textit{Phys. Rev.} \textbf{138}, B979 (1965).
\bibitem{Pegg_1970} D.~T.~Pegg and G.~W.~Series \textit{J. Phys. B: Atom. Molec. Phys.} \textbf{3} L33-5 (1970).
\bibitem{Stenholm_1972} S.~Stenholm, \textit{J. Phys. B: Atom. Molec. Phys.} \textbf{5} 876 (1972).
\bibitem{Cohen-Tannoudji_1973a} C.~Cohen-Tannoudji, J.~Dupont-Roc and C.~Fabre, \textit{J. Phys. B: Atom. Molec. Phys.} \textbf{6} L214 (1973).
\bibitem{Cohen-Tannoudji_1973b} C.~Cohen-Tannoudji, J.~Dupont-Roc and C.~Fabre, \textit{J. Phys. B: Atom. Molec. Phys.}  \textbf{6}, L218 (1973)

\bibitem{Haroche_2016} S.~Haroche and J.~M.~Raimond: \textit{Exploring the Quantum Atoms Cavities, and Photons} (Oxford Graduate Texts 2006).
\bibitem{Girvin_2011} S.~M.~Girvin in \textit{Proceedings of the 2011 Les Houches Summer School on Quantum Machines}, M.~H.~Devoret, R.~J.~Schoelkopf and B.~Huard editor (Oxford University Press 2014).
\bibitem{Leibfried_2003} D.~Leibfried, R.~Blatt, C.~Monroe and D.~Wineland, \textit{Rev. Mod. Phys.} \textbf{75}, 281 (2003).

% Deep strong  coupling regime
\bibitem{Ciuti_2005}C.~Ciuti, G.~Bastard, and I.~Carusotto, \textit{Phys. Rev. B} \textbf{72}, 115303 (2005).
\bibitem{Anappara_2009}A.~A.~Anappara, S.~De~Liberato, A.~Tredicucci, C.~Ciuti, G.~Biasiol, L.~Sorba, and F.~Beltram, \textit{Phys. Rev. B} \textbf{79}, 201303(R) (2009).
\bibitem{Forn_Diaz_2010} P.~Forn-D\'{i}az, J.~Lisenfeld, D.~Marcos, J.~J.~Garc\'{i}a-Ripoll, E.~Solano, C.~J.~P.~M.~Harmans, and J.~E.~Mooij, \textit{Phys. Rev. Lett.} \textbf{105}, 237001 (2010).
%\bibitem{Casanova_2010} J.~Casanova, G.~Romero, I.~Lizuain, J.~J.~Garc\'{i}a-Ripoll, and E.~Solano, \textit{Phys. Rev. Lett.} \textbf{105}, 263603 (2010).
\bibitem{Niemczyk_2010} T.~Niemczyk, F.~Deppe, H.~Huebl, E.~P.~Menzel, F.~Hocke, M.~J.~Schwarz, J.~J.~Garcia-Ripoll, D.~Zueco, T.~H\"{u}mmer, E.~Solano, A.~Marx and R.~Gross, \textit{Nat. Phys.} \textbf{6}, 772 (2010).
\bibitem{Pedernales_2015} J.~S.~Pedernales, I.~Lizuain, S.~Felicetti, G.~Romero, L.~Lamata and E.~Solano, \textit{Sci. Rep.} \textbf{5}, 15472 (2015).
\bibitem{Kraglund_2017} C.~K.~Andersen and A.~Blais, \textit{New J. Phys.} \textbf{19}, 023022 (2017).
\bibitem{Forn-Diaz_2019} P.~Forn-D\'{i}az, L.~Lamata, E.~Rico, J.~Kono, and E.~Solano, \textit{Rev. Mod. Phys.} \textbf{91} 025005 (2019).
\bibitem{Kockum_2019} A.~F.~Kockum, A.~Miranowicz, S.~De~Liberato, S.~Savasta and F.~Nori, \textit{Nat. Rev. Phys.} \textbf{1}, 19 (2019).

% Energy Spectrum
\bibitem{Tomka_2014} M.~Tomka, O.~El~Araby, M.~Pletyukhov, and V.~Gritsev, \textit{Phys. Rev. A} \textbf{90}, 063839 (2014).
\bibitem{Tomka_2015} M.~Tomka, M.~Pletyukhov, and V.~Gritsev, \textit{Sci. Rep.} \textbf{5}, 13097 (2015).
\bibitem{Braak_2011} D.~Braak, \textit{Phys. Rev. Lett.} \textbf{107}, 100401 (2011).

%Emulated Hamiltonians
%\bibitem{Solano_2003} E.~Solano, G.~S.~Agarwal, and H.~Walther, \textit{Phys. Rev. Lett} \textbf{90}, 027903 (2003)
\bibitem{Ballester_2012} D.~Ballester, G.~Romero, J.~J.~Garc\'{i}a-Ripoll, F.~Deppe, and E.~Solano \textit{Phys. Rev. X} \textbf{2}, 021007 (2012)
\bibitem{Dimer_2007} F.~Dimer, B.~Estienne, A.~S.~Parkins, and H.~J.~Carmichael, \textit{Phys. Rev. A} \textbf{75}, 013804 (2007).
\bibitem{Morales_2018} A.~Morales, P.~Zupancic, J.~L\'{e}onard, T.~Esslinger, and T.~Donner, \textit{Nat. Mater} \textbf{17}, 686 (2018).

%\bibitem{Casteels_2017} W.~Casteels, R.~Fazio and C.~Ciuti, \textit{Phys. Rev. A}, \textbf{95}, 012128 (2017).
\bibitem{Gutierrez_2018} R.~Guti\'{e}rrez-J\'{a}uregui and H.~J.~Carmichael, \textit{Phys. Rev. A} \textbf{98}, 023804 (2018).
\bibitem{Kroeze_2018} R.~M.~Kroeze, Y.~Guo, V.~D.~Vaidya, J.~Keeling, and B.~L.~Lev, \textit{Phys. Rev. Lett.} \textbf{121}, 163601 (2018).
%\bibitem{Mavrogordatos_2019} Th.~K.~Mavrogordatos, \textit{Phys. Rev. A} \textbf{100}, 033810 (2019).


\bibitem{Siddiqi_2013} K.~W.~Murch, S.~J.~Weber, K.~M.~Beck, E.~Ginossar and I.~Siddiqi, \textit{Nature} \textbf{499}, 62 (2013).
\bibitem{Siddiqi_2016} D.~M.~Toyli, A.~W.~Eddins, S.~Boutin, S.~Puri, D.~Hover, V.~Bolkhovsky, W.~D.~Oliver, A.~Blais, and I.~Siddiqi, \textit{Phys. Rev. X} \textbf{6}, 031004 (2016).


\bibitem{Huang_2009} S.~Huang and G.~S.~Agarwal, \textit{Phys. Rev. A} \textbf{80}, 033807 (2009).
\bibitem{Agarwal_2016} G.~S.~Agarwal and S.~Huang, \textit{Phys. Rev. A} \textbf{93}, 043844 (2016).
\bibitem{Qin_2018} W.~Qin, A.~Miranowicz, P.~B.~Li, X.-Y.~L\"{u}, J.~Q.~You, and F.~Nori, \textit{Phys. Rev. Lett.} \textbf{120}, 093601 (2018).
\bibitem{Agarwal_2020} C.~J.~Zhu, L.~L.~Ping, Y.~P.~Yang and G.~S.~Agarwal, \textit{Phys. Rev. Lett.} \textbf{124}, 073602 (2020).


\bibitem{Leroux_2018} C.~Leroux, L.~C.~G.~Govia, and A.~A.~Clerk, \textit{Phys. Rev. Lett.} \textbf{120}, 093602 (2018).


\bibitem{Carmichael_2008} H.~J.~Carmichael: \textit{Statistical Methods in Quantum Optics 1 {\&} 2} (Springer-Verlag Berlin, Heidelberg, 1999 and 2008).


\bibitem{Kimble_1986} Ling-An~Wu, H.~J.~Kimble, J.~L.~Hall, and Huifa~Wu, \textit{Phys. Rev. Lett.} \textbf{57}, 2520 (1986).
\bibitem{Kimble_1987} Ling-An~Wu, Min~Xiao, and H.~J.~Kimble \textit{J. Opt. Soc. Am. B} \textbf{4}, 1465 (1987).
\bibitem{Agarwal_libro} G.~S.~Agarwal: \textit{Quantum Optics} (Cambridge University Press, Cambridge, 2012).
% Driven dissipative systems


\bibitem{Lo_1991} C.~F.~Lo, Quantum Opt. \textbf{3}, 333 (1991).




\bibitem{Reid_1986} M.~D.~Reid and D.~F.~Walls,  \textit{Phys. Rev. A} \textbf{34}, 4929 (1986).
%
\bibitem{Corzo_2013} N.~V.~Corzo, Q.~Glorieux, A.~M.~Marino, J.~B.~Clark, R.~T.~Glasser, and P.~D.~Lett, \textit{Phys. Rev. A} \textbf{88}, 043836 (2013).
%
\bibitem{Gutierrez_2018b} R.~Guti\'{e}rrez-J\'{a}uregui and H.~J.~Carmichael, \textit{Phys. Scr.} \textbf{93}, 104001 (2018).


\bibitem{Gilles_1998} L.~Gilles, B.~M.~Garraway, and P.~L.~Knight, \textit{Phys. Rev. A} \textbf{49}, 2785 (1994).




% Cat States and ground state solutions

\bibitem{Agarwal_1993} K.~Tara, G.~S.~Agarwal, and S.~Chaturvedi, \textit{Phys. Rev. A} \textbf{47}, 5024 (1993).

\bibitem{Minganti_2016} F.~Minganti, N.~Bartolo, J.~Lolli, W.~Casteels and C.~Ciuti, \textit{Sci. Rep.} \textbf{6}, 26987 (2016). % In this work it is shown that the ground states of the system remain like cat states for a big part of the parameter regime for a Kerr non-linearity driven by an squeezed drive. The system is similar to ours sans the two level system. Analytic results are obtained.
\bibitem{Ashhab_2010} S.~Ashhab and F.~Nori, \textit{Phys. Rev. A} \textbf{81}, 042311 (2010). 
\bibitem{Maldonado_2019} F.~H.~Maldonado-Villamizar, C.~Huerta~Alderete, and B.~M.~Rodr\'{i}guez-Lara, \textit{Phys. Rev. A} \textbf{100}, 013811 (2019).


\bibitem{Yuen_1976} H.~P.~Yuen, \textit{Phys. Rev. A} \textbf{13}, 2226 (1976).
\bibitem{Mandel_1995} L.~Mandel and E.~Wolf: \textit{Optical coherence and quantum optics} (Cambridge University Press, New York, 1995).

\bibitem{Carmichael_2008b} H.~J.~Carmichael: \textit{Statistical Methods in Quantum Optics 2} (Springer-Verlag Berlin, Heidelberg, 2008), Sec.~12.1.7.

% Multi-photon resonances
\bibitem{Agarwal_1991} G.~S.~Agarwal, \textit{Phys. Rev. A} \textbf{43}, 1523 (1991).
\bibitem{Shamailov_2010} S.~S.~Shamailov, A.~S.~Parkins, M.~J.~Collett, and H.~J.~Carmichael, \textit{Opt. Commun.} \textbf{283}, 766 (2010).
\bibitem{Rempe_2017} C.~Hamsen, K.~N.~Tolazzi, T.~Wilk, and G.~Rempe, \textit{Phys. Rev. Lett.} \textbf{118}, 133604 (2017).

%\bibitem{Minganti_2018} F.~Minganti, A.~Biella, N.~Bartolo, and C.~Ciuti, \textit{Phys. Rev. A} \textbf{98}, 042118 (2018).


\end{thebibliography}
\end{document}